\documentclass[aps,twocolumn,nofootinbib,superscriptaddress]{revtex4-1}
\usepackage{amsmath,amsthm,amsfonts,mathptmx,fullpage}
\usepackage{stmaryrd}
\usepackage{ftnxtra}
\usepackage{tablefootnote}
\usepackage[margin=0.7in]{geometry}

\usepackage[small,bf]{caption}
\usepackage{subcaption}

\usepackage{tablefootnote}

\def\c1{{\ctrl{1} }}
\def\c2{{\ctrl{2} }}
\def\b1{{\ctrl{-1} }}
\def\b2{{\ctrl{-2} }}

\usepackage{algorithm} 
\usepackage{algpseudocode,pifont}

\floatname{algorithm}{Algorithm}

\usepackage{hyperref} 

\usepackage{graphicx,color,cases} 
\usepackage{xfrac}  
\definecolor{DarkGreen}{rgb}{0.1,0.5,0.1}
\definecolor{DarkRed}{rgb}{0.5,0.1,0.1}
\definecolor{DarkBlue}{rgb}{0.1,0.1,0.5} 
 
\newtheorem{theorem}{Theorem} 
\newtheorem{lemma}[theorem]{Lemma}

\theoremstyle{definition}

\def\PI{{\texttt{PI}}}
\def\GNU{\texttt{GNU} }
\def\>{\rangle} 
\def\<{\langle}
 
 \def\case#1{{\left\{  
	\begin{array}{ll}
  #1
	\end{array}
 \right.   }} 

\newcommand{\bi}[2]{\binom{#1}{#2}}

\DeclareMathOperator{\tr}{Tr}  
\def\floor#1{{\left \lfloor {#1} \right \rfloor  }}

\begin{document}

\title{Quantum storage in quantum ferromagnets}

\author{Yingkai Ouyang}\email[y.ouyang@sheffield.ac.uk]
{}
\affiliation{Department of Physics and Astronomy, University of Sheffield, Sheffield, UK}
\affiliation{Singapore University of Technology and Design, 8 Somapah Road, Singapore}
\affiliation{Centre for Quantum Technologies, National University of Singapore, 3 Science Drive 2, Singapore}

\begin{abstract} 
We must protect inherently fragile quantum data to unlock the potential of quantum technologies.
A pertinent concern in schemes for quantum storage is their potential for near-term implementation.
Since Heisenberg ferromagnets are readily available, we investigate their potential for robust quantum storage. 
We propose to use permutation-invariant quantum codes to store quantum data in Heisenberg ferromagnets, because the ground space of any Heisenberg ferromagnet must be symmetric under any permutation of the underlying qubits.
By exploiting an area law on the expected energy of Pauli errors, we show that increasing the effective dimension of Heisenberg ferromagnets can improve the storage lifetime.
When the effective dimension of Heisenberg ferromagnets is maximal, we also obtain an upper bound for the storage error. 
This result relies on perturbation theory, where we use Davis' divided difference representation for Fr{\'e}chet derivatives along with the recursive structure of these divided differences.
Our numerical bounds allow us to better understand how quantum memory lifetimes can be enhanced in Heisenberg ferromagnets.
\end{abstract} 
 
\maketitle

{\em Introduction.---}
Decoherence quickly renders unprotected quantum data unreliable. To combat this, it becomes necessary to encode quantum data into quantum error correction codes.
The challenge in designing robust quantum memories arises from the difficulty of simultaneously (i) utilizing an easily accessible physical system, (ii) having a quantum code that lies within the ground space of the system's Hamiltonian $H$,
and (iii) having an increased storage lifetime $\tau$ with an increasing number of qubits $N$ in the physical system.
Self-correcting quantum memories \cite{RevModPhys.87.307,brown2016quantum} should satisfy (ii) and (iii), but are challenging to implement in a multitude of desirable settings \cite{DAP02,Kit06,KaC08,BrT09,CRL10,CLBT10,Pastawski2011,BrH13,brell2016proposal,PhysRevLett.113.260504,Has11,Yos11}. Indeed, constraint (i) does easily not hold, which frustrates the design of reliable quantum storage. 
For instance, quantum memories based on stabilizer codes which correct at least one error and also satisfy (ii) unfortunately reside in unphysical systems with many-body interactions, and can only be approximately constructed \cite{Kit06,JordanFahri-gadgets,OckoYoshida11,BFBD11}. 
Of these constraints, it is most pertinent to satisfy (i), because physically unrealistic quantum memories will be difficult to engineer.

There are two reasons to store quantum data within the ground space and thereby satisfy constraint (ii).
First, a growing energy gap can suppress excitations from the ground space \cite{marvian2017error}. 
Second, storing quantum data within the ground space avoids unnecessary errors that can occur even in the complete absence of noise. Any state within the ground space is an eigenstate of the physical system, and for such states, they are left unchanged by a unitary operation $U_\tau$ that the system's natural dynamics induces, after a storage time of $\tau$ elapses. By avoiding the need to uncompute $U_\tau$, we would not suffer from an imperfect reversal of $U_\tau$ caused by our imprecise knowledge of $\tau$.

Storage within the ground space, while satisfying constraint (ii), is not enough to result in self-correcting quantum memories and thereby satisfy constraint (iii). 
Moreover, many physically realistic systems satisfying constraint (i) comprising of two-local terms are surprisingly incompatible with constraint (iii) \cite{PhysRevLett.113.260504}. 
However, this no-go result does not preclude physical systems comprising of non-commuting two-body interactions from satisfying constraint (iii).
Consequently, determining whether such physical systems can satisfy constraint (iii) is especially pertinent. 
In this paper, we study Heisenberg ferromagnets as media for quantum storage because they comprise of non-commuting two-body interactions and therefore sidestep the no-go result of \cite{PhysRevLett.113.260504}. 
We also study to what extent Heisenberg ferromagnets satisfy constraint (iii).

The Heisenberg ferromagnet (HF) is a model of quantum magnetism, and is prevalent in many naturally occurring physical systems, and thereby satisfies constraint (i). 
For instance, the HF is found in various cuprates \cite{PhysRevLett.76.3212,chung2001large}, in solid Helium-3 \cite{thouless1965exchange}, and more generally in systems with interacting electrons \cite{Blundell}.
Even in many physical systems that cannot be naturally interpreted as ferromagnets,
effective HFs can nonetheless be engineered, for instance by symmetrizing systems dominated by dipole interactions using dynamic pulse sequences \cite{UhrigPRL}.
Effective HFs have also been engineered in ultracold atomic gases  \cite{duan2003controlling} and quantum dots \cite{tamura2004tunable}.
Specifically, we study spin-half HFs in the absence of an external magnetic field, with Hamiltonian of the form
\begin{align}
\smash{  H  
= -  \sum_{ \{i,j\} \in E } J   ({\bf 1} -\pi_{i,j})} .\label{eq:HH}
\end{align}
Here, ${\bf 1}$ denotes an $N$-qubit identity operator, $\pi_{i,j}$ denotes a swap operator on the $i$th and $j$th qubits, $J$ denotes the exchange constants,
and $E$ denotes the set of interactions.
Such HFs have $J>0$ and ground state energy set to zero.

By storing quantum data using permutation-invariant ({\PI}) codes in HFs, we automatically satisfy constraint (ii). 
This is because symmetric states lie within the ground space of any HF, and quantum data in {\PI} codes, by being invariant under any permutation of their underlying qubits, are symmetric states. 
The possibility of quantum error correction using a {\PI} code was first affirmed by Ruskai, via construction of an explicit {\PI}-$\llbracket 9,1,3\rrbracket$ code that corrects one error \cite{Rus00} \footnote{Here, $\llbracket N,K,D\rrbracket$ denotes a quantum code that encodes $K$ logical qubits into $N$ physical qubits and with distance $D$.
When one needs to correct $t$ errors on a quantum code, it suffices to have $D >  2t$.}.
Ruskai's result \cite{Rus00} was subsequently improved to yield \smash{{\PI}-$\llbracket 7,1,3\rrbracket$} codes \cite{PoR04}.
In \cite{ouyang2014permutation}, Ouyang constructs \smash{{\PI}-$\llbracket (2t+1)^2,1,2t+1\rrbracket$} codes that correct $t$ arbitrary errors, thereby generalizing Ruskai's nine-qubit code.
Further generalizing these codes, Ouyang introduces 
{\PI}-$\smash{\llbracket (2t+1)^2(d-1),\log_2 d,2t+1\rrbracket}$ codes \cite{OUYANG201743}.
Approximate quantum error correcting {{\PI}} codes were also investigated in qubit \cite{ouyang2014permutation,ouyang2015permutation,PhysRevLett.123.110502} and bosonic settings \cite{ouyang2019permutation}.
Research on {\PI} codes shows that quantum correction is possible within the ground space of HFs, 
which is only suggestive that constraint (iii) can be compatible with {\PI} codes.
This is because the coding parameters of {\PI} codes alone, being independent of the parameters in HFs, are not enough to determine what happens when physical noise applies to {\PI} codes stored in HFs.
To better understand the extent in which HFs with {\PI} codes can satisfy constraint (iii), we study bounds on the storage error of {\PI} codes under the action of two different noise models, where both bounds depend on properties of the underlying HF. 

Our first noise model applies to HFs of any geometry, and introduces Pauli errors. These Pauli errors occur with probabilities that are thermodynamically related to their expected energies on the codespace of a specific family of {\PI} codes \cite{ouyang2014permutation}.
To derive an upper bound on the storage error, we find an area law on the expected energy of a Pauli error, which demonstrates that a quantum memory based in a HF can have a macroscopic energy barrier for Pauli errors.
This allows us to show that the storage error decreases with increasing dimensionality of the HF.

Our second noise model introduces unitary errors probabilistically, where each unitary arises from a local perturbation of the underlying Hamiltonian.
Such a noise model can describe the effects of unwanted physical interactions, such as spurious local fields afflicting each particle independently.
We use perturbation theory to bound the storage error by 
using Davis' divided difference representation of these taking Fr{\'e}chet derivatives. 
Because we require complete knowledge of the Hamiltonian's spectrum, 
we restrict our analysis to exactly solvable mean-field HFs. In such HFs every pair of spins interacts with equal strength. Since such HFs have an infinite effective dimension, analyzing them is indicative of the ultimate limits of robust quantum storage in HFs.

With respect to both noise models, we provide upper bounds for the storage error of quantum memories in HFs that are numerically tractable. 
In both cases, we find that quantum memories in HFs are partially self-correcting in the sense that is an optimal system size for fixed noise parameters that minimizes our upper bounds on the storage error.

  {\em Energy of Pauli errors and their geometry.---}
  We use \GNU codes \cite{ouyang2014permutation} to elucidate the dependence of a HF's dimension with respect to the storage error of quantum data.
\GNU codes depend on three parameters $g$, $n$, and $u$, 
and encode a single qubit into $N=gnu$ qubits.
Here $g$ and $n$ quantify the distance of the \GNU code with respect to bit-flip and phase-flip errors respectively, while $u$ is a scaling parameter where $u\ge 1$.
When $g=n=2t+1$, the \GNU code corrects $t$ errors.
A \GNU code has logical codewords 
\begin{align}
|r_L\> =  \sum_{\substack{0 \le j \le n \\ {\rm mod}(j,2)=r} } 
\sqrt{\frac{\binom n j}{ 2^{n-1} } } 
|D^{gnu}_{gj}\>	 ,
\end{align}
where $r=0,1$, and $|D^N_w\>$ are Dicke states of weight $w$ \footnote{Dicke states are uniform superpositions of computational basis states labeled by binary vectors of Hamming weight $w$.} \cite{TGG09,BGu13}.

We quantify a HF's dimension using properties of its underlying graph of interactions \cite{ouyang2017spectrum}.
This graph $G$ has vertices labeled from 1 to $N$, and edges $E$ that correspond to the interactions in the HF's Hamiltonian $H$.
Given a subset $S$ of $\{1,\dots,N\}$, let $\partial_E S$ denote its edge-boundary with respect to the edge set $E$, which is the set of edges in $E$ with exactly one vertex in $S$.
When every subset $S$ satisfies the inequality $|\partial_E S| \ge c \min(|S|,N-|S|)^{1-1/\delta}$, 
the graph and HF have dimension $\delta$ with isoperimetric constant $c$.

\begin{figure}
  \centering
  \includegraphics[width=0.9\linewidth]{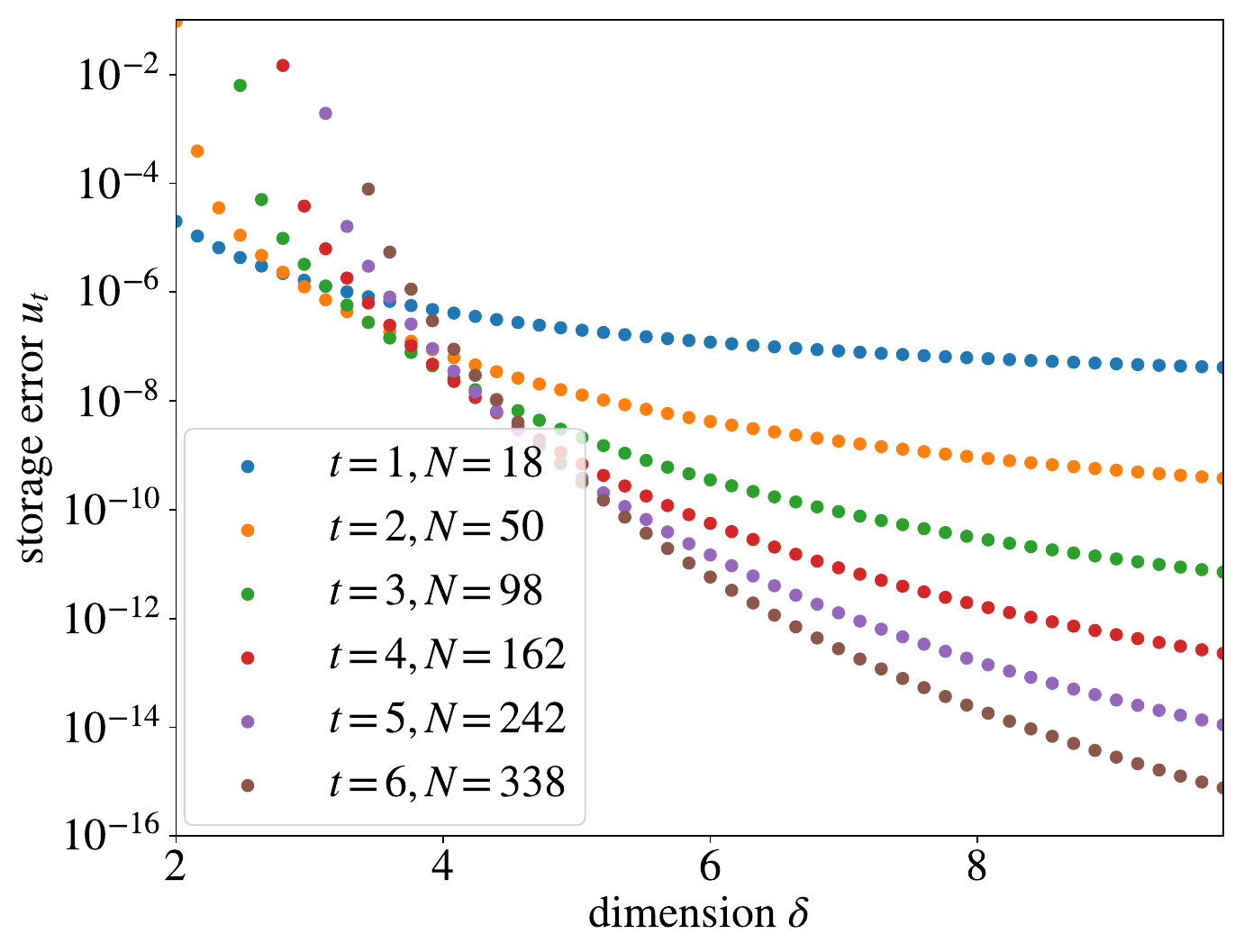}
  \caption{
  When a HF stores an encoded qubit within a \GNU code on $N=2(2t+1)^2$ physical qubits that corrects $t$ errors, we use \eqref{eq:failure-bound} to obtain upper bounds on the storage error $u_{t}$ with respect to the HF's dimension $\delta$ and $t$. Here $\beta J=13$ and $c=1$.}
  \label{fig:geometry}
\end{figure}

Given a set $\mathcal P$ of $N$-qubit Pauli errors that afflict at most $N/2$ qubits, let
\begin{align}
\<P\>  = \min_{|\psi\> \in \mathcal C} 
   \<\psi| P H  P |\psi\>
\end{align} 
denote the minimum expected energy of $P\in \mathcal P$ on the code $\mathcal C$.
When $\mathcal C$ is a \GNU code,
we derive a lower bound on $\<P\>$ in terms of the edge boundary $V(P)$, where $V(P)$ denotes the set of vertices on which $P$ 
acts non-trivially. 
In particular, Theorem \ref{thm:geometric-bound} below gives an area law on the minimum size of $\<P\>$.
\begin{theorem} \label{thm:geometric-bound}
Let $\mathcal C$ be an $N$-qubit \GNU code with parameters $g=n=2t+1$ and $u=2$, where $N=2(2t+1)^2$ and $t \ge 1$.
Then with respect to the Hamiltonian $H$ given by \eqref{eq:HH} with exchange constants $J$ and set of interactions $E$, for every $N$-qubit Pauli $P$ in $\mathcal P$, we have 
\begin{align}
\<P\>\ge \chi  J |\partial_E (V(P))|, \notag
\end{align}
where
$\chi = \min\{2\mu, 1- 4\mu\}$ and
\begin{align}
\smash{\mu = (1+5t+6t^2)/(4+32t+32t^2)}.\notag
\end{align}
\end{theorem}
The significance of Theorem \ref{thm:geometric-bound} 
lies in the geometric interpretation it imparts to $\<P\>$.
Namely, when the graph $G$ has dimension $\delta$ and isoperimetric number $c$, we have the isoperimetric inequality
\begin{align}
\smash{\<P\> \ge J \chi c |P|^{1-1/\delta}},\label{eq:energy-iso-ineq}
\end{align}
where $|P|=|V(P)|$ denotes the weight of $P$.
For a HF on a 1D spin-chain, $\smash{\<P\>\ge J \chi}.$
For a HF on a square lattice, \cite{BoL91} with our result implies that $\smash{\<P\>\ge J \chi \sqrt{|P|}}.$
Whenever $\delta>1$, the expected energy of $P$ grows with its weight, and we have a macroscopic energy barrier \cite{campbell2019theory}. 
This suggests that when $\delta>1$, HFs can be good quantum memories. 
To see this, consider a noise model that introduces Pauli errors $P \in \mathcal P$ with probability proportional to $e^{-\beta \<P\>}$ and with effective inverse temperature $\beta$. 
The corresponding channel is $\mathcal T$, which randomly introduces Pauli errors, and has the form
\begin{align}
\mathcal T(\rho) = \sum_{P \in \mathcal P} (e^{-\beta\<P\>}/ \mathcal Z) P \rho P ,
\label{random-Pauli-channel}
\end{align}
where $\mathcal Z = \sum_{P\in \mathcal P} e^{-\beta \<P\>} $. 
The corresponding probability of obtaining an uncorrectable error, which is also the storage error after perfectly performing quantum error correction, is then 
\begin{align}
u_{t} 
=
\frac{1}{\mathcal Z}\sum_{\substack{ P \in \mathcal P\\ |P| \ge t+1 }} e^{-\beta \<P\>} .
\label{eq:ut-error}
\end{align}
From the isoperimetric inequality \eqref{eq:energy-iso-ineq} and the bound $|\partial_E V(P)| \le  \Delta |P|$, where $\Delta$ is the maximum vertex degree of $G$, we obtain 
\begin{align}
u_{t}
\le 
\left(\sum_{w= t+1 }^{N/2} \binom N w 3^w e^{-\beta J \chi c w^{1-1/\delta}}
\right)
\left(\sum_{w=0 }^{N/2} \binom N w 3^w e^{-\beta J \Delta w}
\right)^{-1}.\label{eq:failure-bound}
\end{align}
We illustrate \eqref{eq:failure-bound} in Figure \ref{fig:geometry} with $\Delta=4, c=1, \beta J = 13$, and vary the dimension $\delta$ and number of correctible errors $t$. 
Figure \ref{fig:geometry} shows that long-range interactions can help to suppress $u_{t}$ because increasing $\delta$ decreases 
$u_{t}$. 
From Figure \ref{fig:geometry}, when $\delta < 4$, the optimal code corrects between 1 error to 4 errors. This shows that for a system with sufficient few dimensions, our scheme exhibits a partial self-correction property, where increasing the system size cannot indefinitely reduce the storage error. 

{\em Random coherent noise and storage error.---}
A good quantum memory preserves entanglement.
Given a quantum code $\mathcal C$ with logical codewords $|0_L\>,\dots, |(M-1)_L\>$, consider the entangled state 
$|\Psi_{\mathcal C}\> =   
\sum_{j=0}^{M-1} |j\>\otimes |j_L\> / \sqrt M.$
The {\em storage error} of $\mathcal C$ with respect to a noisy channel $\mathcal N$ is
\begin{align}
    \epsilon(\mathcal N, \mathcal C)
    = \min_{\mathcal R} \frac{1}{2} \left\|
    |\Psi_{\mathcal C}\>\<\Psi_{\mathcal C}| -
  \overline{\mathcal R} (\overline{\mathcal N} ( |\Psi_{\mathcal C}\>\<\Psi_{\mathcal C}| ) )   \right\|_1,\notag
\end{align}
where $\overline {\mathcal R} = \mathcal I \otimes \mathcal R$, 
$\overline {\mathcal N} = \mathcal I \otimes \mathcal N$,
$\mathcal I$ is an identity channel, 
the minimization is over all recovery maps $\mathcal R$, and $\| \cdot \|_1$ denotes the trace norm.  
To keep notation simple, when the code $\mathcal C$ and noise model $\mathcal N$ are already specified, we use the shorthand $\epsilon = \epsilon(\mathcal N, \mathcal C)$.

\begin{figure}
  \centering
  \includegraphics[width=0.9\linewidth]{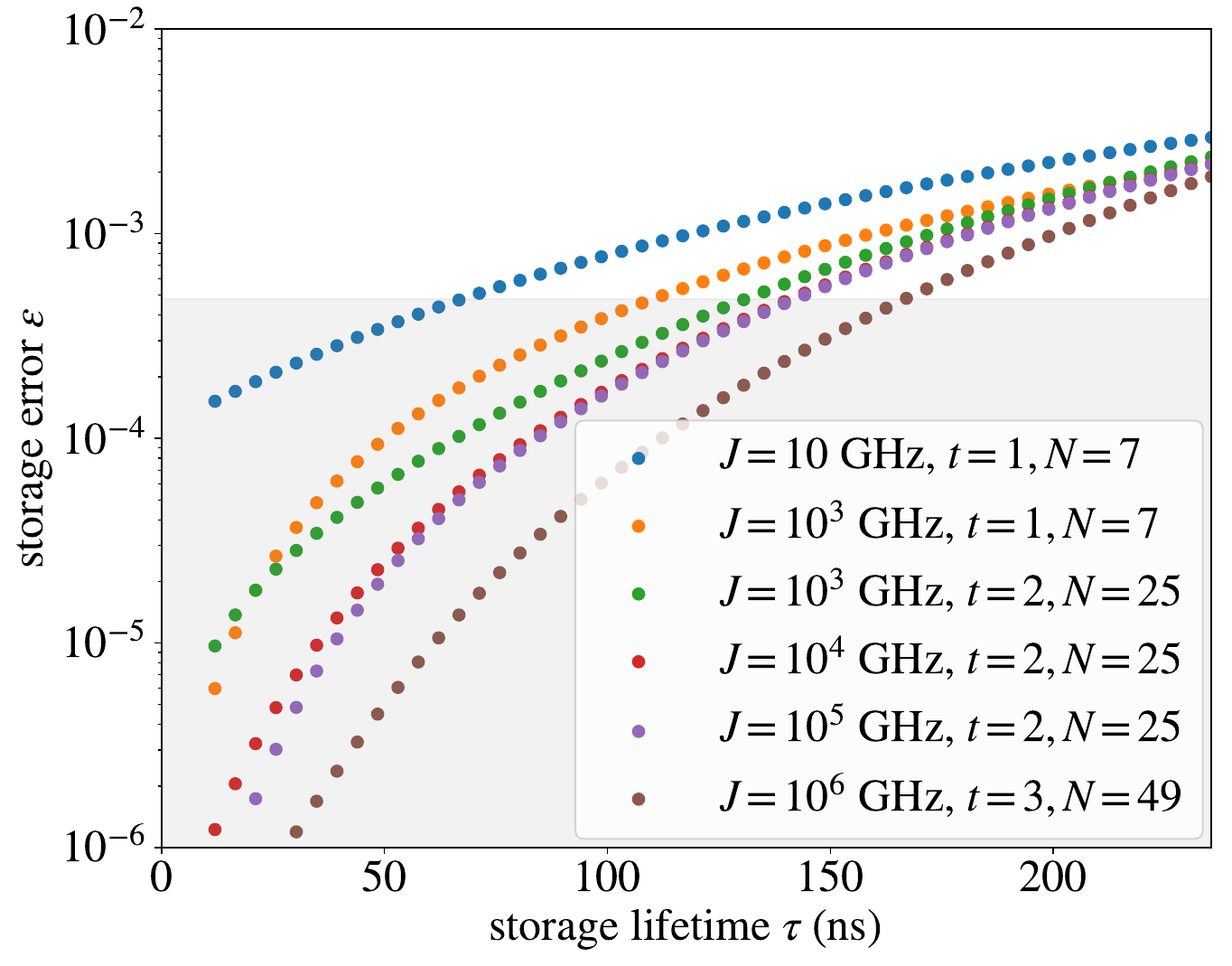}
 \caption{
 When a mean-field HF stores an encoded qubit within an $N$-qubit {\PI} that corrects $t$ errors, we use Theorem \ref{thm:mfmodel} to obtain upper bounds for the corresponding storage error $\epsilon$ after a target storage lifetime of $\tau$.
  The baseline lifetime and storage error for an unprotected qubit are $12$ns and $0.00048$ respectively. The shaded region indicates where $\epsilon$ is smaller than the baseline.}
 \label{fig:sub2}
\end{figure}

Now, let perturbations $A_1,\dots, A_\alpha$ to the Hamiltonian $H$ occur with probabilities $p_1,\dots,p_\alpha$ respectively. 
These perturbations model the effect of qubits coupling to spurious classical fields.
Each perturbation $A_j$ is a linear combination of operators that acts non-trivially on a single qubit,
and induces a unitary evolution $\smash{U_j = g(H+A_j)}$, where $g(x)=e^{-ix\tau}$ denotes an exponential function.
The corresponding noisy channel is $\mathcal N_{\tau}$, where for any initial state $\rho$, we have
\begin{align}
\smash{ \mathcal N_{\tau}(\rho) 
= \sum_{j=1}^\alpha p_j U_j \rho U_j^\dagger}.\label{eq:random-coherent-noise}
\end{align}  
In what follows, we focus our attention solely $\mathcal N_{\tau}$, which we call a random coherent noise channel, which is parameterized by its noise strength $\smash{a = \max_j \| A_j\|/ N}$.

For any perturbation $A_j$, the Taylor series of the unitary $g(H+A_j)$ gives
\begin{align}
\smash{g(H+A_j) = g(H) + \sum_{k = 1}^\infty  D^{[k]}_g (H,A_j)/ k!},
\end{align}
where
\begin{align}
\smash{ D^{[k]}_g (H,A_j)
= 
{ \frac{d^k}{d\xi^k} } g(H+\xi A_j) |_{\xi=0} }
\end{align}
 are Fr{\'e}chet derivatives of $g(H)$ in the matrix direction $A_j$ \cite{bhatia2013matrix,deadman2016-frechet-remainder}.
Now the correctible component of $\smash{g(H+A_j)}$
with respect to a code that corrects $t$ errors comprises of Fr{\'e}chet derivatives of order at most $t$, 
because these Fr{\'e}chet derivatives are polynomials in $A_j$ of order no more than $t$.
Therefore, we study only the high order Fr{\'e}chet derivatives. These Fr{\'e}chet derivatives allow us to bound the storage error.
\begin{lemma}\label{lem:error-Rbound}
Given a quantum code $\mathcal C$ that corrects $t$ errors, 
let $\smash{R_j = \sum_{k=t+1}^\infty D^{[k]}_g (H,A_j)}$ and define 
$\|R\|_{\mathcal C} = \max_j \{\|R_j|\psi\>\|: |\psi\>\in \mathcal C\}$.
Then 
$\epsilon   \le  \|R\|_{\mathcal C} + \|R\|_{\mathcal C}^2.$
\end{lemma}
From the integral representation of $R_j$ \cite{ouyang2019compilation}, we exploit the fact that $g(H+A_j)$ is unitary for Hermitian $H$ and $A$ to get 
\begin{align}
\|R\|_{\mathcal C}  \le  \max_j \frac{\max\{\|D^{[t+1]}_g (H,A_j)|\psi\> \|: |\psi\> \in \mathcal C\}}{(t+1)!},
\label{eq:first-R-bound}
\end{align}
which depends only on a single Fr{\'e}chet derivative instead of infinitely many. 
Given \eqref{eq:first-R-bound}, one can clearly bound $\|R\|_{\mathcal C}$ in terms of just $\|H\|$ and $aN$. 
However, such a bound increases with increasing $\|H\|$, and exhibits a behavior contrary to the numerical evidence in Figure \ref{fig:geometry}. 
Increasing the number of long-range interactions increases both $\|H\|$ and dimensionality, 
but since increasing dimensionality should decrease the storage error,
this suggests that the storage error should instead decrease with increasing $\|H\|$.
We solve this conundrum by using Davis' representation \cite{davis1973LAA-divided-diff} of Fr{\'e}chet derivatives, which reveals the intricate dependence of Fr{\'e}chet derivatives on the spectral decomposition $H = \sum_{j\ge 0} \lambda_j \Pi_j$. Here, $\lambda_j$ strictly increase with $j$ and $\Pi_j$ are eigenprojectors. 
Namely, we can write $\smash{D^{[k]}_g (H,A_j)/k!}$ as
\begin{align}
 \sum_{n_0,\dots,n_k} 
g(\lambda_{n_0},\dots ,\lambda_{n_k}) 
( \Pi_{n_k} A_j) \dots ( \Pi_{n_1} A_j ) \Pi_{n_0} .
\label{eq:frechet-divdiff}
\end{align}
Here, $g(\lambda_{n_0},\dots ,\lambda_{n_k}) $ are divided differences that arise naturally from the theory of Lagrange interpolation.
To unravel \eqref{eq:frechet-divdiff}, we leverage on the remarkable properties of divided differences. 
First, divided differences are invariant under any permutation of their arguments.
Hence, we can always arrange the arguments of a divided difference in non-decreasing order.
Second, divided differences generalize scalar derivatives, because the divided difference of a vector with $k$ identical arguments is proportional to the $(k-1)$th derivative of the underlying function. For the exponential function, we have $|g(y_1, \dots, y_k)|  =   \tau^{k-1}/(k-1)!$ when 
$y_1 = \dots = y_k$.
For instance, $|g(2,2)| \le \tau $.
Third, a divided difference when not evaluated on identical arguments can be recursively defined; whenever $y_i$ and $y_j$ are distinct, 
$  \smash{  g({\bf y}) = 
    ( g({\bf y}[{\rm not}\ i]) - g({\bf y}[{\rm not}\ j]) )/(    y_i  -  y_j )},$
where ${\bf y}[{\rm not}\ i]$ denotes a vector obtained from ${\bf y}$ by deleting its $i$th component.
From \eqref{eq:first-R-bound} and \eqref{eq:frechet-divdiff},
we find that 
\begin{align}
\| R\|_{\mathcal C} 
\le  (aN)^{t+1} \left(h_{t+1} + \frac{ \tau^{t+1} }{ (t+1)! } \right).
\label{eq:remainder-bound}
\end{align}
Here $ \tau^{t+1} /(t+1)! $ arises from the divided difference with all arguments equal to zero.
The term $h_{t+1}$ is the sum of all $|g(0,\lambda_{n_1},\dots ,\lambda_{n_{t+1}})|$ where $n_1+ \dots+ n_{t+1}>0$.

Evaluating a bound on $h_{t+1}$ requires knowing the eigenvalues of $H$.
Since finding the eigenvalues of $H$ for arbitrary $E$ is in general a difficult problem \cite{ouyang2017spectrum}, we study an exactly solvable HF where every pair of spins interacts equally with $J_{i,j} = J$. We call such a HF a mean-field HF, and its ground state energy is $\lambda_0 = 0$ and its higher energy eigenvalues are
$\lambda_1=J N, \lambda_2 = 2J(N-1)$, $\lambda_3 = 3J(N-2)$,
 and in general, $\lambda_j = J j(N+1-j) $ \cite{ouyang2017spectrum}.
We proceed to bound $\delta_j$, which is the minimum energy needed to transition away from $\lambda_j$.
For instance, $\delta_0 = \lambda_1- \lambda_0 = JN $, 
$\delta_1 = \lambda_2-\lambda_1 = J(N-2)$,
and $\delta_2 = \lambda_3-\lambda_2 = J(N-4)$. 
In general, $\delta_{\floor{N/2}} = 2 + (N-2\floor{N/2})$ and 
$\delta_j = J(N-2j)$ for all $j=0,\dots, \floor{N/2}-1$.
Importantly, $\delta_j$ is non-increasing in $j$ and is maximal when $j=0$.
Exploiting the recursive structure of divided differences, one can show that 
\begin{align}
|g(0,\lambda_{n_1},\dots ,\lambda_{n_t})| 
\le 2^{t+1}\delta_0^{-1} 
(\delta_{n_1}\dots \delta_{n_{t+1}} )^{-1} .
\label{eq:distinct-divided-diff-eig-bound}
\end{align}
When there are repeated arguments for the divided difference, the bound is more complicated than \eqref{eq:distinct-divided-diff-eig-bound}, because it involves maximization.
To avoid maximizing, we overestimate the size of $h_{t+1}$ by overcounting the contributions from 
$g(0,\lambda_{n_1},\dots ,\lambda_{n_{t+1}})$ when there are repeated arguments in $g$.
First we use \eqref{eq:distinct-divided-diff-eig-bound} for divided differences even when there are $r$ repeated arguments.
Second, for divided arguments with $r$ repeated entries, we count the contributions from leaves that terminate with all possible divided differences with repeated identical arguments. 

If the contribution to the divided differences is dominated by leaves with no repeating indices, the total contribution of such leaves to $h_{t+1}$ is 
at most $\smash{S^{t} 2^{t+1}/\delta_0}$, where 
$\smash{S = \delta_0^{-1} + \dots + \delta_{t+1}^{-1}}$.
The contribution to $h_{t+1}$ by leaves that terminate with $r$ repeated arguments is $\smash{(\tau^{r-1}/(r-1)!) S^{t+1-r}/\delta_0}$.
From this we get $h_{t+1}\le \theta$ where
\begin{align}
\theta =  \frac{S^{t+1}}{\delta_0}
+ \frac{n/2+1}{\delta_0}\sum_{r=2}^{t+1} 
\frac{\binom{t+1}{r-1} \tau^{r-1}}{(r-1)!}S^{t+2-r} .
\label{eq:h-coarse-bound}
\end{align}
This leads us to our final result, which follows directly from Lemma \ref{lem:error-Rbound}, \eqref{eq:remainder-bound} and \eqref{eq:h-coarse-bound}.
  \begin{theorem}\label{thm:mfmodel}
Let $H$ be a mean-field HF where every pair of qubits interacts with exchange constant $J$, and $\mathcal C$ be any {\PI} code that corrects $t$ errors. 
 Let $\mathcal N_\tau$ be the random coherent noise channel as given in \eqref{eq:random-coherent-noise}. 
  Then when we have perfect quantum error correction, 
  $\epsilon(\mathcal N_\tau,\mathcal C)
\le \Theta + \Theta^2$, where $\smash{\Theta = (aN)^{t+1}(\theta +  \tau^{t+1}/(t+1)!)}$,
and $\theta$ is given in \eqref{eq:h-coarse-bound}.
  \end{theorem}
  From Theorem \ref{thm:mfmodel}, we can see that the quantum memory is partially self-correcting, because the bound on the storage error contains a term $(N \tau)^t/t!$ which diverges for large $N$, since $N$ is quadratic in $t$. Hence for every noise parameter $a$ and exchange constant $J$, our scheme for a quantum memory has an optimal system size. Figure \ref{fig:sub2} illustrates only results for optimal system sizes.
  
   Recently, a superconducting qubit was stored between 12ns to 20ns with a fidelity of 0.9995 \cite{Barends2014}. 
Using our noise model, these experimental parameters can be recast into a baseline storage error of $5(10^{-4})$ with a memory lifetime of $12$ns and a noise strength of $a = 0.04$MHz. 
Given this noise model, we use Theorem \ref{thm:mfmodel} to obtain upper bounds on the storage error $\epsilon$ of an encoded qubit within a {\PI} code in a mean-field HF, and we depict these numerical results in Figure \ref{fig:sub2}. Here, the number of qubits for $t=1$ is seven \cite{PoR04}, and when $t\ge 2$, $N=(2t+1)^2$. 
From Figure \ref{fig:sub2}, if one uses a seven-qubit {\PI} code with $J=10^3$GHz, the qubit's storage lifetime can be improved to over 100ns.
In addition, if one uses a 25 qubit {\PI} code that corrects 2 errors \cite{ouyang2014permutation}, the qubit's storage lifetime can be enhanced to over 120ns when $J=10^4$\textsc{GHZ}. 
Similarly, if $J=10^6$\textsc{GHZ}, the qubit's storage lifetime can be enhanced to over 150ns using a 49 qubit {\PI} code that corrects 3 errors.
From this, we can see how increasing $J$ and the number of qubits in HFs can enhance the storage lifetime.

{\em Discussions.---}
Here, we study quantum storage in a physically abundant physical system, the HF.
Our scheme, being based on a physical model that is simple to realize, will be easier to implement than those built upon many-body interactions.
Because Pauli errors on {\PI} codes can exhibit a macroscopic energy barrier,
we see evidence that a quantum memory based in a HF can become increasingly robust with increasing dimensionality of the HF. 
In addition, we study the impact of increasing the coupling strengths within HFs on the targeted storage error. 
For this, we analyze an infinite-dimensional HF, namely the mean-field HF, and find numerically tractable upper bounds on the storage error.
Our derivation of the bounds relies on a novel approach based on the connection between matrix perturbation theory and divided differences. 
We find that increasing the size of the coupling strengths can be beneficial in extending the storage lifetime of a HF-based quantum memory when used in concert with a {\PI} code.

Since our analysis technique extends to any physical system with a completely understood spectral structure, it applies also to other code-inspired Hamiltonians, and lays the foundations for analyzing using quantum memories using our new methodology.
While {\PI} codes can be robustly prepared in superconducting charge qubits, by inducing Rabi-like oscillations between Dicke states of adjacent weights \cite{init-picode-2019-PRA}, it remains to see how the initialization Hamiltonian can be integrated with the HF.
Furthermore, constructing explicit protocols for the decoding of {\PI} codes can bring quantum memories in HFs closer to implementation.

\section{Acknowledgments}
{\em Acknowledgements.---} YO thanks Joseph F. Fitzsimons for coming up with the idea for studying Heisenberg ferromagnets for quantum storage, and also for helping to prove Theorem \ref{thm:geometric-bound}. YO also thanks Jon Tyson for bringing to his attention the divided difference representation of matrix Fr\'echet derivatives. YO acknowledges support from the Singapore National Research Foundation under NRF Award NRF-NRFF2013-01, the U.S. Air Force Office of Scientific Research under AOARD grant FA2386-18-1-4003. YO acknowledges support from the EPSRC (Grant No. EP/M024261/1) and the QCDA project (Grant No. EP/R043825/1) which has received funding from the QuantERA ERANET Cofund in Quantum Technologies implemented within the European Union’s Horizon 2020 Programme.
\bibliography{mems}{}
\bibliographystyle{ieeetr}

\appendix

\section{Proof of Theorem \ref{thm:geometric-bound}}
 We now show that the larger the weight of $P$, the larger the expected energy of $P \rho P$, where $\rho$ is supported on our chosen {\PI} codespace. 
By definition of $\<P\>$, we have 
\begin{align*}
\<  P \>
& = -\sum_{\{i,j\} \in E} J  ( \tr(  P {  \rho}   P {  \pi}_{i,j} ) - \tr({  \rho} {  \pi}_{i,j}) ). 
\end{align*}
Since ${  \rho}$ is permutation-invariant, both ${  \pi}_{i,j} {  \rho}$ and ${  \rho} {  \pi}_{i,j}$ are equal to ${  \rho}$, and hence 
\begin{align}
\<P\> =J \sum_{\{i,j\}\in E}   T_{i,j}, 
\end{align}
where 
\begin{align}
T_{i,j} = 1-\tr({  \rho} ({  \pi}_{i,j}   P {  \pi}_{i,j})   P ). 
\end{align}
If swapping the $i$th and $j$th qubits leaves $P$ invariant, we have $({  \pi}_{i,j}   P {  \pi}_{i,j})   P = I ^{\otimes N}$, and the corresponding term $T_{i,j}$ is zero. 
Now denote $X_i$, $Y_i$ and $Z_i$ as Pauli operators that apply a single $X$, $Y$ and $Z$ Pauli on the $i$th qubit and leave the remaining qubits untouched.

Conversely, if swapping the $i$th and $j$th spins changes $P$, the operator $({  \pi}_{i,j}   P {  \pi}_{i,j})   P$ is either 
$ X_{i,j}$, $Y_{i,j}$, or $Z_{i,j}$. 
Because $\rho$ is permutation-invariant, it suffices to evaluate 
\begin{align}
\tr ({  \rho} X_{1,2} ), \quad
\tr ({  \rho} Y_{1,2} ), \quad
\tr ({  \rho} Z_{1,2} ).
\end{align}
Since $\rho$ is supported on a {\PI}-code that corrects at least a single error, the Knill-Laflamme conditions \cite{KnL97} imply that for all $Q \in \mathcal Q = \{X_{1,2}, Y_{1,2}, Z_{1,2}\}$,
we have $\< 0_L | Q | 1_L\>= \< 1_L | Q | 0_L\>=  0$, and $\< 0_L | Q | 0_L \> = \< 1_L | Q | 1_L \>$ \cite{ouyang2014permutation}.
Hence for all $ Q \in \mathcal Q$ and $\rho$ supported on our {\PI}-codespace, 
\begin{align*}
\tr( \rho Q) 
&= 
\<0_L|  \rho |0_L\>\< 0_L | Q | 0_L \> 
+
\<1_L|  \rho |1_L\>\< 1_L | Q | 1_L \> \notag\\
&= 
\< 0_L |Q | 0_L \> 
=\frac{ \< 0_L | Q | 0_L \> + \< 1_L | Q | 1_L \>}{2}
\notag\\
&=
\sum_{ 	\substack { 
	0 \le \ell \le n \\
}} 
2^{-n} \bi{n}{\ell} \< D^m_{g\ell} | Q | D^m_{g \ell } \>  . 
\end{align*}
Using the Vandermonde binomial identity 
$  \bi{N}{g\ell} =  \bi{N-2}{g\ell} + 2\bi{N-2}{g\ell-1} + \bi{N-2}{g\ell-2}$, 
it follows that 
\begin{align}
\tr({  \rho}   Q ) = \case{ 
	1- 4 \mu,	 & Q = Z \\
	2 \mu, & Q \in\{X,Y\} \\
	} ,
\end{align}
 where for our \GNU code with parameters $g,n$ and $u$, we have
\[
\mu := \sum_{0 \le \ell \le n} 2^{-n} \bi{n}{\ell} \frac{ \bi{N-2}{ g\ell -1 } }{\bi{N}{g\ell}}. 
\]
In particular, for all $u > 1$, we have $0 < \mu < 1/2$, which implies that either $T_{i,j} = 0$ or $T_{i,j}$ is strictly positive. The lower bound for $\<P\>$ then depends crucially on the number of edges for which $T_{i,j}$ is strictly positive.
This implies that
\begin{align}
\<  P \>
& \ge  J |\partial_E(V(P))| \min\{1-4\mu, 2 \mu \}. 
\end{align}
Substituting $u=2, g=n=2t+1$ then gives the result.

\section{Proof of Lemma \ref{lem:error-Rbound}}
Consider $\mathcal N(\rho) = \sum_j p_j U_j \rho U_j^\dagger$ for any density matrix $\rho$, where $p_j$ are probabilities and $U_j$ are unitary matrices. 
Let $\mathcal C$ be a quantum code with logical codewords $|r_L\>$ for $r=0,\dots, d-1$. 
Let $U_j= G_j + B_j$, where $G_j$ and $B_j$ denote the correctible and uncorrectable parts of $U_j$ respectively with respect to $\mathcal C$.
It suffices to show that 
\begin{align}
\epsilon(\mathcal N, \mathcal C)  
 \le 
 \sum_{j}  \frac{p_j }{d}
   \sum_{r=0}^{d-1}
\left( \sqrt{\<r_L|  B^\dagger  B |r_L\>}
+  \<r_L|  B^\dagger  B |r_L\> \right) .\label{lem2:showthis}
\end{align}
We now proceed with the proof.
From the linearity of the channels $\mathcal N$ and $\mathcal R$, it is clear that
\begin{align}
    \epsilon(\mathcal N, \mathcal C)
    =& 
    \min_{\mathcal R} \frac{1}{2} \left\|
    |\Psi_{\mathcal C}\>\<\Psi_{\mathcal C}| -
  \overline{\mathcal R} (\overline{\mathcal N} ( |\Psi_{\mathcal C}\>\<\Psi_{\mathcal C}| ) )   \right\|_1
  \notag\\
    =& 
    \min_{\mathcal R} \frac{1}{2} \left\|
   \frac{1}{d} \sum_{r,s=0}^{d-1}
\left(    |r\>\<s| \otimes |r_L\>\<s_L| 
\right.\right.\notag\\
&-
\left.\left.
    |r\>\<s| \otimes \sum_{j} p_j \mathcal R\left( U_j |r_L\>\<s_L| U_j ^\dagger 
  \right)
  \right)    \right\|_1  .\notag
\end{align}
Using the triangle inequality for the trace norm and the monotonicity of the trace distance under the partial trace, we get
\begin{align}
\epsilon(\mathcal N, \mathcal C)  
 \le 
 \sum_{j}  \frac{p_j }{2 d}
   \sum_{r=0}^{d-1}
\left\| M_{r,U_j}\right\|_1,
\end{align}
where
\[M_{r,U} = |r_L\>\<r_L|  - \mathcal R ( U|r_L\>\<r_L| U^\dagger).\]
It remains to obtain an upper bound on $\|M_{r,U}\|_1$.
\begin{lemma} \label{lem:tracenorm-MrU}
\begin{align}
\|M_{r,U}\|_1 \le 2 \sqrt{\<r_L|  B^\dagger  B |r_L\>}
+ 2 \<r_L|  B^\dagger  B |r_L\>.
\end{align}
\end{lemma}
We defer the proof of Lemma \ref{lem:tracenorm-MrU}, which essentially involves decomposing $M_{r,U}$ into a block matrix on the correctible and uncorrectable subspaces of the quantum code $\mathcal C$.
Thus, it readily follows that
\[
\epsilon(\mathcal N, \mathcal C)  
 \le 
 \sum_{j}  \frac{p_j }{d}
   \sum_{r=0}^{d-1}
\left( \sqrt{\<r_L|  B^\dagger  B |r_L\>}
+  \<r_L|  B^\dagger  B |r_L\> \right) .
\]
Since probabilities sum to one, we can use H\"older's inequality and find that the maximization over $r$ gives the upper bound required in the lemma.

\section{Proof of Lemma \ref{lem:tracenorm-MrU}}
Let $\Pi_g$ and $\Pi_b$ denote projectors onto the correctible and uncorrectable subspaces of the quantum code $\mathcal C$ respectively.
It follows that $\Pi_g |r_L\> = |r_L\>$ and $\Pi_b |r_L\> = 0$.
Hence, by writing $M_{r,U}$ into a size 2 block matrix induced by the projectors $\Pi_g$ and $\Pi_b$ we get
\begin{align} 
M_{r,U}= 
\begin{pmatrix}
m_{g,g} & m_{g,b} \\
m_{b,g} & m_{b,b}\\
\end{pmatrix},
\end{align}
where
\begin{align}
m_{g,g} &=
|r_L\>\<r_L| - \Pi_g \mathcal R ( U|r_L\>\<r_L|  U^\dagger)  \Pi_g ,
\notag\\
m_{g,b} &=  \Pi_g  \mathcal R ( U|r_L\>\<r_L|  U^\dagger)  \Pi_b,
\notag\\
m_{b,g} &=  \Pi_b  \mathcal R ( U|r_L\>\<r_L|  U^\dagger)  \Pi_g ,
\notag\\
m_{b,b} &=   \Pi_b  \mathcal R ( U|r_L\>\<r_L|  U^\dagger)  \Pi_b.
\end{align}
Let $U= G+B$ where $G$ denotes the correctible part of $U$ and $B$ denotes the uncorrectable part of $U$ with respect to the code $\mathcal C$.
Second, since $\mathcal R$ corrects errors on the correctible subspace and its Kraus operators do not induce interactions between the correctible and uncorrectable subspaces,
it follows that
\begin{align}
&\Pi_g \mathcal R ( U|r_L\>\<r_L|  U^\dagger)  \Pi_g \notag\\
&=
\Pi_g \mathcal R ( G |r_L\>\<r_L|  G^\dagger)  \Pi_g 
+\Pi_g \mathcal R ( B |r_L\>\<r_L|  G^\dagger)  \Pi_g \notag\\
&+\Pi_g \mathcal R ( G |r_L\>\<r_L|  B^\dagger)  \Pi_g 
+\Pi_g \mathcal R ( B |r_L\>\<r_L|  B^\dagger)  \Pi_g \notag\\
&=
 |r_L\>\<r_L|  \< r_L |G^\dagger G |r_L \> .
\end{align}
 Since $\<r_L |U^\dagger U|r_L\>$ =1, it follows from the definition of $G$ and $B$ that $\<r_L| G^\dagger G |r_L\> + \<r_L| B^\dagger B |r_L\> = 1$. Hence
 \begin{align}
M_{r,U} =
\begin{pmatrix}
\<r_L| B^\dagger B|r_L\> |r_L\>\<r_L| 
&  \Pi_g  \mathcal R ( U|r_L\>\<r_L|  B^\dagger)  \Pi_b\\
  \Pi_b  \mathcal R ( B |r_L\>\<r_L|  U^\dagger)  \Pi_g &   \Pi_b  \mathcal R ( B|r_L\>\<r_L|  B^\dagger)  \Pi_b
\end{pmatrix}.
\end{align}
We now require a block-matrix generalization of the Ger\u sgorin circle theorem \cite{varga-GCT} given by the following lemma.
\begin{lemma} \label{lem:block-gct}
Let $M$ be a Hermitian matrix with the block decomposition 
\begin{align}
M=  \begin{pmatrix}
A & C \\ C^\dagger & B \\
\end{pmatrix}, \notag
\end{align}
where $A$ and $B$ are Hermitian matrices of the same size.
Then 
\begin{align}
\|M\|_1 \le \|A\|_1 +\|B\|_1 + 2\|C\|_1.
\end{align}
\end{lemma}
\begin{proof}
Since $M$ is Hermitian, there is a unitary matrix $V$ such that $\|M\|_1 =\tr (VM)$. Such a unitary matrix $V$ admits a block decomposition
\begin{align}
V =  
\begin{pmatrix}
V_{11} & V_{12} \\ V_{21} & V_{22} \\
\end{pmatrix}.
\end{align}
Then we can see that 
\begin{align}
\|M\|_1  = \tr(V_{11} A  + V_{12} C + V_{21} C^\dagger + V_{22} B). \label{eq:traces-abc}
\end{align}
Since the trace of the product of two matrices is an inner product, we can use H\"older's inequality to obtain an upper bound on $\|M\|_1. $
Such a bound requires knowledge on the operator norm (the maximum singular value) of the submatrices of $V$.
For this, we use the parallelogram law. Note that
\begin{align}
V \begin{pmatrix}
{\bf x} \\ {\bf y} \\
\end{pmatrix}
&=
\begin{pmatrix}
V_{11} {\bf x} + V_{12} {\bf y} \\  V_{21} {\bf x} + V_{22} {\bf y} \\
\end{pmatrix}
\notag\\
V \begin{pmatrix}
{\bf x} \\ -{\bf y} \\
\end{pmatrix}
&=
\begin{pmatrix}
V_{11} {\bf x} - V_{12} {\bf y} \\  V_{21} {\bf x} - V_{22} {\bf y} \\
\end{pmatrix}.
\end{align}
 Whenever $({\bf x} ; {\bf y})$ is a vector of unit norm, it follows that the vectors on the right side of the above equation are also vectors of unit norms.
 Now let us apply the parallelogram law, which gives us
 \begin{align}
2\|V_{11} {\bf x}  \|^2 + 2\|V_{12} {\bf y}  \|^2 
= \|V_{11} {\bf x} + V_{12} {\bf y}  \|^2   
+ \|V_{11} {\bf x} - V_{12} {\bf y}  \|^2  .
\end{align}
Since $V$ is unitary and $({\bf x} ; {\bf y})$ is a vector of unit norm, we always have
 \begin{align}
2\|V_{11} {\bf x}  \|^2 + 2\|V_{12} {\bf y}  \|^2 
\le 2  .
\end{align}
Whenever ${\bf y}=0$, we must have $\|{\bf x}\| = 1$, and in this scenario we find that $\|V_{11} {\bf x}  \|^2 \le 1$.
This implies that $\|V_{11}\| \le 1$. Repeating this argument, we find that the operator norm of all submatrices of $V$ is at most 1. 
Hence the result follows from applying H\"older's inequality on \eqref{eq:traces-abc}.
\end{proof}
Using Lemma \ref{lem:block-gct}, we get
\begin{align}
&\|M_{r,U} \|_1 \le \<r_L| B^\dagger B|r_L\> 
+ 2 \| \Pi_g  \mathcal R ( U|r_L\>\<r_L|  B^\dagger)  \Pi_b \|_1 \notag\\
 &+ \|\Pi_b  \mathcal R ( B|r_L\>\<r_L|  B^\dagger)  \Pi_b\|_1.
\end{align}
Now recall that for matrices $S$ and $T$, $\|ST\|_1 \le \|S\| \|T\|_1$
and $\|ST\|_1 \le \|T\| \|S\|_1$.
Using this fact repeatedly, we find that
\begin{align}
 \| \Pi_g  \mathcal R ( U|r_L\>\<r_L|  B^\dagger)  \Pi_b \|_1 
 &\le 
  \| \mathcal R ( U|r_L\>\<r_L|  B^\dagger)  \|_1 ,\\
 \|\Pi_b  \mathcal R ( B|r_L\>\<r_L|  B^\dagger)  \Pi_b\|_1
 &\le  
  \| \mathcal R ( B|r_L\>\<r_L|  B^\dagger) \|_1.
\end{align}
Furthermore, because $\mathcal R$ is a trace-preserving map, we have
\begin{align}
 \| \Pi_g  \mathcal R ( U|r_L\>\<r_L|  B^\dagger)  \Pi_b \|_1 
 &\le 
  \|  |r_L\>\<r_L|  B^\dagger  \|_1 ,\\
 \|\Pi_b  \mathcal R ( B|r_L\>\<r_L|  B^\dagger)  \Pi_b\|_1
 &\le  
  \|  B|r_L\>\<r_L|  B^\dagger \|_1.
\end{align}
Using that fact that for a matrix $S$, we have $\|S\|_1 = \tr \sqrt{S^\dagger S} $, it follows that
\begin{align}
 \|  |r_L\>\<r_L|  B^\dagger  \|_1 
 &= \tr \sqrt{B |r_L\>\<r_L|  B^\dagger } \notag\\
 &=  \sqrt{\<r_L|  B^\dagger  B |r_L\>},
\end{align}
and
\begin{align}
  \|  B|r_L\>\<r_L|  B^\dagger \|_1
  &=
  \tr\sqrt{ B|r_L\>\<r_L|  B^\dagger  B|r_L\>\<r_L|  B^\dagger }\notag\\
  &=\<r_L|  B^\dagger  B |r_L\>.
\end{align}
Hence 
\begin{align}
\|M_{r,U} \|_1  
&\le 2 \sqrt{\<r_L|  B^\dagger  B |r_L\>}
+ 2 \<r_L|  B^\dagger  B |r_L\>,
\end{align}  
and the result follows.

 \section{Proof of Eq.~\eqref{eq:distinct-divided-diff-eig-bound}}
  
\begin{figure}
\centering
\includegraphics[width=0.4\textwidth]{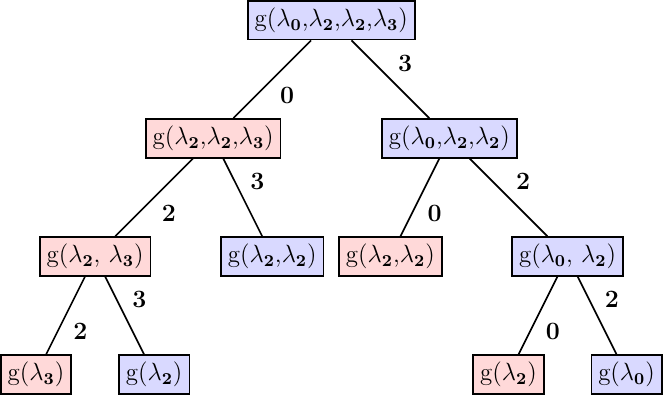}
\caption{A tree illustrates the recursive structure of the divided difference of $g(\lambda_0,\lambda_2,\lambda_2, \lambda_3)$, where $g$ is the exponential function.}
\label{fig:tree}
\end{figure}

In Figure \ref{fig:tree}, we depict $g(\lambda_0,\lambda_2,\lambda_2,\lambda_3)$ as a binary tree with root labeled by $g(\lambda_0,\lambda_2,\lambda_2,\lambda_3)$.
Children of the root are obtained by removing the largest and the smallest arguments from $g(\lambda_0,\lambda_2,\lambda_2,\lambda_3)$, which are 
$\lambda_0$ and $\lambda_3$ respectively.
Hence, the root's children inherit a factor of 
$(\lambda_3-\lambda_0)^{-1}$ from the root.
This procedure iterates until no distinct arguments in the divided difference remain.
We can see that the leftmost vertex in Figure \ref{fig:tree} is a leaf labeled by $g(\lambda_3)$ that contributes at most $(\lambda_3-\lambda_0)^{-1}(\lambda_3-\lambda_2)^{-2}$ to the bound. 
The blue vertex labeled by $g(\lambda_2,\lambda_2)$ contributes at most 
$(\lambda_3-\lambda_0)^{-1}(\lambda_3-\lambda_2)^{-1} \tau$ to the bound. 
Using the triangle inequality and the monotonicity of $\delta_i$, we find that   
$g(\lambda_0,\lambda_2,\lambda_2,\lambda_3)\le 
\max\{ 
2^{3}\delta_0^{-1} \delta_2^{-1}\delta_3^{-1}, 
2^{2}\delta_0^{-1}\delta_3^{-1} \tau 
\}.$
Here, the factor of $\delta_0^{-1}$ arises as a coarse upper bound to $(\lambda_3- \lambda_0)^{-1}$, while $\delta_2$ and $\delta_3$ correspond to the smallest energy gaps obtainable when descending the tree from the root's children onwards.
Generalizing this argument for distinct non-zero 
$\lambda_{n_1},\dots ,\lambda_{n_{t+1}}$ where $n_1 >  \dots > n_{t+1}$
gives the result.
 
\end{document}